\documentclass[twocolumn,showpacs,amsmath,amssymb,aps,pra]{revtex4}
\begin{document}

\title{INFERRING LINEAR QUANTUM DYNAMICS WITHOUT NO-SIGNALING}
\author{S. Gheorghiu-Svirschevski\footnotemark[1]\footnotetext{e-mail: hnmg@soa.com}}
\address{1087 Beacon St., Suite 301, Newton, Massachusetts 02459}
\date{\today}

\begin{abstract}
It is shown that the no-signaling condition is not needed in order to argue a linear quantum dynamics from standard quantum statics and the usual interpretation of quantum measurement outcomes as probabilistic mixtures. 
\end{abstract}


\maketitle

\section{Introduction}

It was claimed recently \cite{Gisin} that the usual "quantum statics" supplemented by the no-signaling condition implies a completely positive, linear quantum dynamical law. The key point of the argument is that the no-signaling condition for spatially separated, noninteracting quantum systems is observed if and only if reduced states and remotely prepared probabilistic mixtures that are statistically equivalent evolve identically. This dynamical equivalence of reduced states and probabilistic mixtures leads then necessarily to dynamical linearity.

We wish to remark however that the dynamical equivalence of mixed states and probabilistic mixtures, and thus a linear quantum dynamics, can be argued under identical assumptions without recourse to the no-signaling condition and space-like quantum nonlocality. The no-signaling condition casts the fundamental statement that "entanglement between noninteracting systems cannot be detected locally" as a corollary of the relativistic limit on the speed of propagation of interactions. Yet an equivalent statement applies to arbitrary sets of noninteracting degrees of freedom, including sets that cannot be associated with spatially separable collections of physical particles. In the latter case, the restriction on entanglement detection is not conditioned by the existence of a relativistic speed limit, but is inherent to the notion of "noninteracting quantum degrees of freedom". We take advantage of quantum "nonlocality" between such spatially non-separable freedoms to show that the no-signaling condition is not necessary in a rationalization of quantum dynamical linearity.  

The fact that the concept of noninteracting freedoms takes precedence over relativistic no-signaling should be no surprise, since no no-signaling argument can be given without reference to noninteracting systems and degrees of freedom. This  implies that an operational definition of the latter is a necessary prerequisite in this context. In other words, a no-signaling argument {\it can} be used to justify a linear dynamics, but it provides only a particular, albeit powerful, realization of a more general inference.

\section{No-signaling, probabilistic mixtures and the linearity of quantum dynamics}
\label{Sec2}
Let us begin by recalling the no-signaling argument for a linear dynamics \cite{Gisin}. Consider an arbitrary quantum system, prepared in a mixed quantum state represented by the {\it probabilistic mixture} $\left \{ {p_i, |\psi_i\rangle} \right \}$. If the density matrix corresponding to this mixture reads 
\begin{equation}
\label{eq1}
\hat \rho = \sum\limits_i {p_i |\psi_i \rangle \langle \psi_i | }\;,
\end{equation}
\noindent a dynamical map $\hat\rho \to {\bf g}(\hat\rho)$, not necessarily linear, decomposes as a superposition of individual evolutions of its pure state components $|\psi_i\rangle$. In other words,
\begin{equation}
\label{eq2}
{\bf g}(\hat\rho) \equiv {\bf g}\left({ \sum\limits_i {p_i |\psi_i \rangle \langle \psi_i | } }\right) = \sum\limits_i {p_i\; {\bf g} \left({ |\psi_i\rangle \langle \psi_i | }\right) }\;,
\end{equation} 
\noindent where ${\bf g} \left({ |\psi_i\rangle \langle \psi_i | }\right)$ need not be a pure state. If the above identity can be applied equally to all probabilistic mixtures described by the density matrix $\hat \rho$, it can be concluded that the dynamical map ${\bf g}$ must be linear. 

The relativistic no-signaling condition is brought in for the purpose of validating the latter conjecture and the first identity in Eq.(\ref{eq2}). Suppose the mixed state described by $\hat\rho$ is a reduced state corresponding to an entangled pure state of space-like separated, noninteracting systems. Such a state can be converted into any equivalent probabilistic mixture by an appropriate maximal measurement on the remote entangled counterpart. But if the no-signaling condition must apply, the local reduced state must evolve in the same manner regardless of its particular realization as a probabilistic mixture, or the remote measurement could be detected locally. Hence, it is argued, the no-signaling condition enforces the first identity in Eq.(\ref{eq2}), and therefore a linear dynamics. 

\section{Linear quantum dynamics without the no-signaling condition}
\label{Sec3}
Nevertheless, the above recourse to the no-signaling condition is not necessary in this context. For let us recall and examine in closer detail the meaning of "noninteracting systems" in a quantum setting. In a traditional understanding, it is said that two distinct systems $\cal A$ and $\cal B$ are noninteracting if and only if interactions restricted to one of them do not affect the time-dependence of any observables pertaining to the other. This is a very general, metatheoretical statement that encompasses both instantaneous and delayed influences, but makes no reference to, and no use of, the presence or absence of a relativistic speed limit for the propagation of interactions. As such, it holds true equally in both nonrelativistic and relativistic physical theories. It also applies {\it a priori }both to uncorrelated systems and classically correlated statistical systems, and thence necessarily to quantum systems in separable probabilistic mixtures [mixed states]. When this definition is further extended to quantum systems in entangled states, it implies that entanglement between noninteracting systems cannot be detected by [local] actions limited to either the observed element or any noninteracting counterpart. To be formally precise, let an entangled state of the systems $\cal A$ and $\cal B$ be described by the common density matrix $\hat\rho_{\cal A + \cal B}(t)$, corresponding to reduced states $\hat\rho_{\cal A}(t)=Tr_{\cal B}\hat\rho_{\cal A+\cal B}(t)$ and, respectively, $\hat\rho_{\cal B}(t)=Tr_{\cal A}\hat\rho_{\cal A + \cal B}(t)$. Then $\cal A$ and $\cal B$ are noninteracting if and only if interactions that involve, say, only observables of $\cal A$, do not affect the time-dependence of the state of $\cal B$, specifically the time-dependence of its ensemble statistics. That is, if ${\bf g}_{\cal A} (\hat\rho_{\cal A + \cal B}(t))$ denotes any new dynamics of $\cal A + \cal B$ induced by interactions involving $\cal A$ only, then $\cal A$ and $\cal B$ are noninteracting if and only if the state of $\cal B$ remains described by the same reduced density matrix, $\hat\rho'_{\cal B}(t)= Tr_{\cal B}{\bf g}_{\cal A}(\hat\rho_{\cal A + \cal B}(t)) = Tr_{\cal A}\hat\rho_{\cal A + \cal B}(t) = \hat\rho_{\cal B}(t)$. 

But this implies, in particular, that interactions leading to a measurement on $\cal A$ at time $t_0$ do not affect the statistics and the density matrix of $\cal B$ at any later time $t > t_0$. Therefore one has from the very definition of "noninteracting systems" that a measurement on one of two such quantum systems that are eventually in an entangled state, whether pure or mixed, must leave the statistics of the other unchanged throughout all later times. If quantum measurements are understood to produce probabilistic mixtures, then it follows again that a suitable choice of measurements on $\cal A$ can prepare the noninteracting counterpart $\cal B$ in any desired probabilistic mixture, while leaving its density matrix $\hat\rho_{\cal B}$ unchanged. Hence probabilistic mixtures and reduced states with identical density matrices must evolve identically, and the overall dynamics must necessarily be linear. 

The potentially questionable point in the line of reasoning sketched above concerns the extension of the definition for "noninteracting systems" over entangled quantum states. It is possible to object that 

\noindent a) in so doing one covertly and necessarily sanctions the relativistic no-signaling condition, or 

\noindent b) the no-signaling condition can always be brought in as a superseding first principle to justify an assumption [specifically, the extension to entangled states] that lacks a more fundamental basis in the existing framework. 

Suppose objection (a) is valid and the described extension to entangled states is justified only if the no-signaling condition is observed, or in effect implies the no-signaling condition. It follows, by logical negation, that if a finite relativistic speed limit is absent, this extension cannot be observed. Yet it is known to be a perfectly consistent and functional notion in the frame of nonrelativistic quantum theory. Hence objection (a) is not pertinent. To clarify objection (b), let us recall that the concept of "noninteracting systems" is superseded in turn by the more fundamental notion of "noninteracting degrees of freedom". The entire discussion above applies identically if the phrase "set of independent degrees of freedom" is substituted for the less descriptive concept of "system". However, unless the individual freedoms are associated with specific space-time locations, as for quantum fields, the notions of space-like separation and nonlocal communication between two degrees of freedom are meaningless. Hence in such cases it is clear that the no-signaling condition cannot be used in the characterization of "noninteracting degrees of freedom". We can exploit this predicament to give a more elaborate version of the previous counterargument, showing that objection (b) can indeed be averted. 

So let us recall that, without any loss of generality, every quantum system is described by a specified set ${\cal S}$ of independent degrees of freedom, represented by commuting coordinates [observables] $q_\alpha$, the allowed values of which label an orthogonal basis in the total Hilbert space of pure quantum states $\cal H_{\cal S}$. Moreover, a total Hilbert space parameterized by a given number of degrees of freedom is isomorphic to a direct product of Hilbert spaces ${\cal H}_\alpha$, each associated with a single  freedom $q_\alpha$. That is, ${\cal H_{\cal S}} \sim \mathop \bigotimes \limits_{q_\alpha \in \cal S}{{\cal H}_\alpha}$. In such a representation, an overall state $\hat\rho$ \cite{Gleason}, whether pure or mixed, can be regarded in general as an entangled state of the independent degrees of freedom \cite{gen-entg}. For every subset of such freedoms ${\cal A} \subset {\cal S}$ one can define a reduced statistics described by a reduced density matrix $\rho_{\cal A} = Tr_{\overline {\cal A}}[ \rho ]$, where ${\overline {\cal A}} = \cal S / \cal A$ denotes the complementary subset of freedoms in the total set ${\cal S}$. On the other hand, according to the usual interpretation of quantum measurements, a measurement involving only a restricted subset of coordinates $\overline {\cal A}$ prepares the complementary subset ${\cal A}$ in a probabilistic mixture. Since observables associated with distinct $q_\alpha$-s commute, the density matrix of this mixture is necessarily identical to the reduced density matrix $\hat{\rho}_{\cal A}$ of subset ${\cal A}$ prior to the measurement. For instance, let a 3-dimensional particle be described by Cartesian components of its momentum, say $p_x$, $p_y$, and $p_z$, and let it be prepared in a pure state $|\Psi\rangle$ given  as [momentum representation]

\[
|\Psi\rangle = \int dp_x \int dp_y \int dp_z \Psi(p_x,p_y,p_z) |p_x, p_y, p_z\rangle \;\;.
\]

\noindent Then an ensemble measurement of $p_x$ will prepare the complementary components $p_y$ and $p_z$ in a probabilistic mixture with a density matrix 

\[
\hat{\rho}_{p_y p_z}=\int dp_x \langle p_x|\Psi\rangle \langle \Psi | p_x \rangle\;\;.
\]\\

Also, let us recall again that two independent quantum degrees of freedom $q_\alpha$ and $q_\beta$ are said to be noninteracting throughout a given evolution if and only if interactions involving $q_\beta$, but not $q_\alpha$, do not affect the time-dependence of any observables pertaining to $q_\alpha$, and similarly for interactions involving $q_\alpha$, but not $q_\beta$. In particular, if $q_\alpha$ and $q_\beta$ are noninteracting freedoms, an ensemble measurement of $q_\alpha$ at time $t_0$ will not affect the ensemble statistics of $q_\beta$ at any later time $t > t_0$.  

We can establish now that, under the above assumptions, the dynamics of any noninteracting subset $\cal A$ of degrees of freedom must be linear, irrespective of the total dimensionality of the system. From this it follows by extension that quantum dynamics must be linear overall. The argument is again similar to the no-signaling argument, although the no-signaling condition itself is not involved, either directly or indirectly. To make the latter point clear, let us illustrate first with a simple example. A detailed account of the general proof follows.

Imagine the 3-dimensional particle mentioned above as being a charged, spin-0 particle, and let us partition its degrees of freedom into two disjoint sets, ${\cal A} = \{p_y, p_z \}$, and $\overline{\cal A} = \{ p_x \}$. Let this particle be subject to an external interaction that couples to the subset $\{p_y, p_z \}$, but not to  $\{ p_x \}$. For instance, let it move in a magnetic field oriented along the $x$ direction. At the initial time $t_0$ the total state of the particle is an "entangled" state of its $x$, $y$ and $z$ degrees of freedom, and the statistics of the $p_y$ and $p_z$ components of the momentum is described by the reduced density matrix $\rho_{p_y p_z}$. Let us assume that in the absence of any other external intervention this reduced density matrix evolves as

\[
\rho_{p_y p_z} \to {\bf g}_t(\rho_{p_y p_z})\;\;.
\]

\noindent  But suppose now that at time $t_0$ the particle undergoes an ensemble measurement involving the $p_x$ component of the momentum and/or its higher moments, including the spatial coordinate $x$. Such a measurement prepares $p_y$ and $p_z$ in a statistically equivalent mixture, say $\sum\limits_{p_y,p_z}{\lambda_{p_y,p_z} |\phi_{p_y,p_z}\rangle \langle \phi_{p_y,p_z}|}$, which evolves necessarily according to a law of the form

\[
\sum\limits_{p_y,p_z}{\lambda_{p_y,p_z} |\phi_{p_y,p_z}\rangle \langle \phi_{p_y,p_z}|} \to \sum\limits_{p_y,p_z}{\lambda_{p_y,p_z} {\bf g}_t(|\phi_{p_y,p_z}\rangle \langle \phi_{p_y,p_z}|)}\;\;.
\]

\noindent Since the $x$ degree of freedom remains decoupled from the $y$ and $z$ degrees of freedom throughout the propagation, by definition the $x$-measurement cannot affect the reduced statistics of $p_y$ and $p_z$ at either $t_0$ or at any later time $t$. It follows that the evolved mixture and the evolved reduced state must be statistically equivalent, hence must have identical density matrices,

\[
{\bf g}_t(\sum\limits_{p_y,p_z}{\lambda_{p_y,p_z} |\phi_{p_y,p_z}\rangle \langle \phi_{p_y,p_z}|}) = \sum\limits_{p_y,p_z}{\lambda_{p_y,p_z} {\bf g}_t(|\phi_{p_y,p_z}\rangle \langle \phi_{p_y,p_z}|})\;\;.
\]

\noindent The above identity holds for any probabilistic mixture $\sum\limits_{p_y,p_z}{\lambda_{p_y,p_z} |\phi_{p_y,p_z}\rangle \langle \phi_{p_y,p_z}|}$, and it can be shown that any such mixture can be generated by a suitably designed $x$-measurement \cite{Gisin}. Hence we must resolve that the corresponding reduced dynamics along $y$ and $z$ must be linear. 

Observe further that the level of generality of this conclusion extends in fact beyond the above one particle example. Since the dynamics described by the map ${\bf g}_t$ is otherwise arbitrary, a similar statement applies on any isomorphic state space, under isomorphic physical conditions. For instance, the reduced Hilbert space subtended by $p_y$ and $p_z$ for the charged 3-dimensional particle is isomorphic to the state space of a spinless particle moving in a 2-dimensional plane, thus the dynamics of a 2-dimensional particle must be linear. Similarly, the same reduced space parameterized by $p_y$ and $p_z$ is isomorphic to the state space of two distinguishable, spinless 1-dimensional particles, hence a two-particle dynamics in a 1-dimensional space must be linear. In particular, the dynamics of two noninteracting, but entangled 1-dimensional particles [eventually localized on distinct linear segments] must be linear. We note that, indeed, no reference to the no-signaling condition was necessary in order to infer a linear quantum dynamics under the given assumptions about quantum measurements and probabilistic mixtures.

Let us return now to the general case and conclude the argument. Let the chosen quantum system be arranged such that the degrees of freedom selected for its description are partitioned into two mutually noninteracting sets, say $\cal A$ and $\overline{\cal A}$, and let ${\cal H}_{\cal A}$ [${\cal H}_{\overline{\cal A}}$] denote the reduced Hilbert space subtended by the degrees of freedom in set $\cal A$ [$\overline{\cal A}$]. We note that such a partition can always be defined so that neither set of freedoms corresponds to a collection of physical particles, and in such a case the space-like separation of the two sets is physically meaningless. Let this system be subjected to an external interaction that couples to the freedoms in subset $\cal A$, but not to those in the complementary set $\overline{\cal A}$. Suppose that at time $t_0$ the system is in a state with density matrix $\rho(t_0)$, such that the reduced statistics of the set $\cal A$ is described by the reduced density matrix $\rho_{\cal A}(t_0) = Tr_{\overline{\cal A}}[\rho(t_0)]$, and let the evolution of this reduced statistics for times $t > t_0$ be described, in the absence of any other actions on the system, by the dynamical map 

\begin{equation}
\label{eq3}
\rho_{\cal A}(t_0) \to \rho_{\cal A}(t) = {\bf g}_t (\rho_{\cal A})\;\;.
\end{equation}

\noindent The original no-signaling argument can now be easily transposed to the present setting, up to the no-signaling condition itself. If at time $t_0$ the system undergoes an ensemble measurement of some coordinates belonging to the set ${\overline{\cal A}}$, the set $\cal A$ is prepared in a probabilistic mixture, say 

\begin{equation}
\label{eq4}
\rho'_{\cal A}(t_0) = \sum\limits_i {p_i\;|\phi^{({\cal A})}_i \rangle \langle \phi^{({\cal A})}_i | }\;.
\end{equation}

\noindent As in the no-signaling argument \cite{Gisin}, any desired probabilistic mixture of $\cal A$ can be prepared by an appropriate choice of the ${\overline{\cal A}}$-measurement. The evolution of such a probabilistic mixture necessarily decomposes as a superposition of individual evolutions of its pure state components $|\phi^{({\cal A})}_i \rangle \in {\cal H}_{\cal A}$, that is 

\begin{equation}
\label{eq5}
\rho'_{\cal A}(t_0) \to \rho'_{\cal A}(t) = \sum\limits_i {p_i \;{\bf g}_t \left({ |\phi^{({\cal A})}_i \rangle \langle \phi^{({\cal A})}_i  | }\right) }\;.
\end{equation} 

\noindent We recall that ${\bf g}_t \left({ |\phi^{({\cal A})}_i \rangle \langle \phi^{({\cal A})}_i | }\right)$ need not be a pure state [e.g., the ${\cal A}$-degrees of freedom may become entangled with the environment]. 

Now, since the $\cal A$ and  $\overline{\cal A}$ degrees of freedom are mutually independent and noninteracting, by definition the $\overline{\cal A}$-measurement cannot affect and cannot distill entanglement in the reduced statistics of the set $\cal  A$ at either $t_0$ or $t$. This clause takes the place of the original no-signaling condition and suffices to guarantee a similar conclusion. Thus one must necessarily have $\rho_{\cal A}(t_0) = \rho'_{\cal A}(t_0)$ and $\rho_{\cal A}(t) = \rho'_{\cal A}(t)$. But this means that

\[
{\bf g}_t (\rho_{\cal A}) = {\bf g}_t (\sum\limits_i {p_i\; |\phi^{({\cal A})}_i \rangle \langle \phi^{({\cal A})}_i | })
\]

\begin{equation}
\label{eq6}
 = \sum\limits_i {p_i\; {\bf g}_t \left({ |\phi^{({\cal A})}_i \rangle \langle \phi^{({\cal A})}_i  | }\right) }
\end{equation}

\noindent for any probabilistic mixture of $\cal A$ with an initial density matrix $\hat\rho_{\cal A}$. Hence any probabilistic mixture of $\cal A$ must evolve linearly, and the map ${\bf g}_t$ is generally linear on the convex set of density matrices $\rho_{\cal A}$. Since ${\bf g}_t$ is otherwise arbitrary, a similar statement extends to any state space isomorphic to ${\cal H}_{\cal A}$. Moreover, since both the system $\cal S$ and the set $\cal A$ are also arbitrary, it follows that dynamical linearity can be inferred for any quantum system by means of an isomorphism with a subset of degrees of freedom for some properly selected larger system, i.e. by means of a generalized embedding.\\

\section{Conclusions}
\label{Sec4}

In essence, both the argument proposed here and the no-signaling argument for quantum dynamical linearity provide a first principles proof of the well known fact \cite{nlin-ftl-1,nlin-ftl-2,nlin-ftl-3,Polchinski,Mielnik} that nonlinear evolution laws are incompatible with quantum probabilistic mixtures and the phenomenon of remote preparation. Nevertheless, it must be kept in mind, as first pointed out in ref.\cite{Bona}, that the concept of remote preparation entails an implicit acceptance of the projection postulate, which is extraneous to quantum statics. For suppose that two noninteracting systems, or sets of degrees of freedom, $\cal A$ and $\cal B$ are initially in an entangled pure state $|\Psi_{\cal A \cal B}\rangle$. Their respective local states are necessarily elementary reduced states that cannot be prepared locally. Let a maximal preparatory measurement be effected and recorded on system $\cal B$ at time $t_0$ as observed in a given reference frame, and let system $\cal A$ be subject to an observation, also immediately recorded, at an arbitrarily close time $t_0+\delta t$. If the preparation leaves either $\cal A$ or $\cal B$ in a probabilistic mixture of pure states, it follows that the joint state of $\cal A$ and $\cal B$ at $t_0+\delta t$ must be a separable mixture, realizable as a statistical superposition of products of local states [otherwise individual measurement outcomes cannot be pure states in either system]. In other words, the total state must undergo a {\it disentanglement process}. But if this total separable state is indeed a probabilistic mixture, then by the very definition of the latter every pair of $\cal A$ and $\cal B$ [${\rm + measuring \; device}$] in an ensemble representative of this mixture must actually exist in one of the contributing product states, say $\hat\rho_A^{out} \otimes\hat\rho_B^{out}$. Let the joint recorded observations of $\cal A$ and $\cal B$ corroborate one such product state in each run of the experiment. This evidently implies that each particular sample of $\cal A$ and $\cal B$ must undergo an evolution from the initial, pre-measurement state $|\Psi_{\cal A \cal B}\rangle$ at $t_0$ to some final, post-measurement state $\hat\rho_{\cal A}^{out} \otimes\hat\rho_{\cal B}^{out}$ at $t_0 + \delta t$, occurring with a certain probability among a set of possible product states. System $\cal A$, in particular, must evolve from a local state $\hat\rho_{\cal A}^{in} = Tr_{\cal B} \left[ |\Psi_{\cal A \cal B}\rangle\langle \Psi_{\cal A \cal B}| \right ]$ at $t_0$ into the local state $\hat\rho_{\cal A}^{out}$ at $t_0+\delta t$. But this is precisely a "projection at-a-distance" process, which must be assumed to occur instantaneously, since $\delta t$ can be made arbitrarily small. Hence the statement {\it ' local measurements produce local and/or remote probabilistic mixtures ' }is seen to be equivalent to the projection postulate.\\

The original motivation behind the no-signaling argument for dynamical linearity was the observation that any attempt "to modify quantum physics, e.g., by introducing nonlinear evolution laws for pure states, ...easily leads to the possibility of superluminal communication" [\cite{Gisin}, pg.1]. From this it was reasoned that, perhaps, the converse may be true, and linearity is secured by the relativistic impossibility of superluminal signaling, in the form of the no-signaling condition. We showed, however, that the relativistic speed limit does not play an essential role in enforcing dynamical linearity. Linear dynamics derives strictly from the probabilistic mixture interpretation of quantum measurements, in conjunction with quantum nonlocality [remote preparation]. Perhaps it may be argued that the Hilbert space structure of the quantum state space is determined by the relativistic structure of space-time \cite{Svetlichny}. But once this structure is accepted as the fundamental component of {\it quantum statics}, and is supplemented by the probabilistic mixture point of view, there is no need for a recurring reference to relativity in order to justify a linear dynamics.


\begin{references}
\bibitem{Gleason} There is no loss of generality in the assumption that the quantum state of the system is described by a density matrix, since the existence of the latter is always guaranteed by Gleason's theorem, given a Hilbert space of pure states and the trace rule for event probabilities.
\end{references}
\end{document}